\documentclass[12pt,preprint]{aastex}
\usepackage{epsfig}
\begin{document}

\title{Southern Galactic Plane Survey Measurements of the Spatial 
Power Spectrum of Interstellar H{\tt I} in the Inner Galaxy}

\author{John M. Dickey, N. M. McClure-Griffiths}
\affil{Dept. of Astronomy, University of Minnesota}
\author{Sne\u{z}ana Stanimirovi\'c}
\affil{Arecibo Observatory, National Astronomy and Ionosphere Center}
\author{B. M. Gaensler\footnote{Hubble Fellow}}
\affil{Center for Space Research, Massachusetts Institute of Technology}
\author{A. J. Green}
\affil{School of Physics, Sydney University}

\begin{abstract}
Using data from the Southern Galactic Plane Survey we have measured the
spatial power spectrum of the interstellar neutral atomic hydrogen
in the fourth Galactic quadrant.
This function shows the same power law behavior that has been 
found for H{\tt I} in the second quadrant of the Milky Way and
in the Magellanic Clouds, with the same
slope.  When we average over velocity intervals broader than
the typical small-scale velocity dispersion, we find that
the slope steepens, from $\simeq -3$ to $\simeq -4$ for
the warm gas, as predicted by
theories of interstellar turbulence if the column density
fluctuations are dominated by variations in the gas density
on small spatial scales.
The cool gas shows a different increase of slope, that
suggests that it is in the regime of turbulence dominated
by fluctuations in the velocity field.
Overall, these results confirm that the small scale
structure and motions in the neutral atomic
medium are well described by a turbulent cascade of
kinetic energy.
\end{abstract}


\section{Introduction}

The structure of the density and velocity fields in the interstellar
medium (ISM) has been studied using a variety of tracers, from
spectral lines in emission and absorption to pulsar scintillation
and Faraday rotation.  A particularly convenient tracer is the
21-cm line of H{\tt I} because it is very widespread and easily 
detectable from both dense and diffuse clouds and from the
``intercloud'' or warm neutral medium.  Two sorts of structure
are traced with this line:  deterministic and stochastic.
The former are coherent features in the density and/or velocity
fields that can be interpreted dynamically as evolving
physical structures, like shells, bubbles, chimneys, clouds,
or streams.  The latter are apparently random variations in
the density of the gas as a function of position and velocity,
that can best be characterized statistically. The two classes
of structure are related.  For example,
as the shells blown by stellar winds and supernova remnants
expand, they ultimately lose their separate identity and 
blend into the background of apparently random density
fluctuations.  This paper is a study of the structure of
the neutral atomic medium in an area of the fourth quadrant
of the Galactic plane mapped in the 21-cm line as part of
the Southern Galactic Plane Survey (SGPS).  Preceeding papers
in this series (McClure-Griffiths et al.  
\markcite{nao1}2000, \markcite{nao2}2001, hereafter
Paper 1 and Paper 2) presented
some apparently deterministic H{\tt I} structures (shells and chimneys)
from the survey; this paper deals with the random fluctuations.

A favorite statistical characterization of the properties of
a random density or velocity distribution is the structure 
function (Kaplan \markcite{kaplan}1966, Rickett \markcite{rickett}1976).
For the ionized medium
this can be measured over a huge range of scales using the
propagation of pulsar signals.  This analysis
leads to ``the big power law in the sky'' (Spangler \markcite{spangler}2001), 
a power law that corresponds to a Kolmogorov cascade of
interstellar turbulence (Cordes and Rickett \markcite{cordes1}1998 and references
therein).  A different function for describing the structure,
which is particularly easy to apply to aperture synthesis 
observations of the brightness of a spectral line, is the
spatial power spectrum. This is the Fourier transform of
the structure function or the magnitude squared of the 
Fourier transform of the brightness distribution (see Goldman
\markcite{goldman}2000, and references therein).  

Early studies of the spatial power spectrum (hereafter SPS) of Galactic H{\tt I}
(Crovisier and Dickey \markcite{crovisier1}1983; Kalberla and Stenholm
\markcite{kalberla}1983)
revealed, quite unexpectedly, that
it has a smooth, power law form with logarithmic slope
-3$\pm$0.2.  This result remained unexplained for 
many years, even after it was demonstrated by Green \markcite{green}(1993) that
the power law shows almost exactly the same slope in
another direction in the outer Galaxy.  More recently
Stanimirovi\'c et al. \markcite{stanimirovic1}(1999) have shown that even in the 
Small Magellanic Cloud the same power law slope is obtained.
The LMC shows similar behavior (Elmegreen et al. \markcite{elmegreen1}2001), as
do Galactic molecular clouds (Zielinsky and St\"utzki \markcite{zielinsky}1999,
Dickman et al. \markcite{dickman}1990, Gautier et al.
\markcite{gautier}1992).  In recent years, the
relationship between this power law and the Kolmogorov
energy cascade is becoming more clear (Lazarian and
Pogosyan \markcite{lazarian1}2000, and references therein, Goldman
\markcite{goldman}2000).
More profoundly, the structure of the ISM as measured either
by the structure function or by the SPS 
can be interpreted as a manifestation of the fractal geometry
of the medium (Elmegreen et al. \markcite{elmegreen1}2001,
Stanimirovi\'c et al. \markcite{stanimirovic1}1999,
Westpfahl et al. \markcite{westpfahl}1999, St\"utzki et al. 
\markcite{stutzki}1998).  

In this paper we concentrate on testing a simple prediction of
the mathematical theory of turbulence pointed out by 
Lazarian and Pogosyan \markcite{lazarian1}(2000).
The theory states that the slope
of the power law of the observed (two dimensional)
SPS should steepen when the region
sampled changes from thin to thick along the line of sight.
%{\bf if} the emissivity fluctuations are dominated by density
%fluctuations in the gas, rather than small scale 
%fluctuations in the velocity field.
``Thick'' and ``thin'' here depend on whether the line of
sight thickness of the gas sampled by a given velocity
bandwidth is more or less than the thermal width of 
the line divided by the velocity gradient along the
line of sight due to differential Galactic rotation.
In the inner Galaxy, the steep velocity gradient allows us to
make the transition from thin to thick slices
by averaging over progressively broader velocity widths.  
The result we find is in good agreement
with the prediction.  This gives strong support
to the Kolmogorov paradigm as an explanation for the power
law, and it indirectly supports the interpretation of 
Elmegreen et al. \markcite{elmegreen1}(2001) that
the H{\tt I} disk in the LMC is thin. A similar
result is found for the LMC by Padoan \markcite{padoan}et
al. (2001) using the related spectral correlation function technique.

The observed steepening of the power law slope of the
SPS with broader velocity averaging
is so clear that we attempt a more profound analysis.
Interstellar turbulence includes both velocity fluctuations
(small scale structure in the velocity field of the gas)
and density fluctuations, each with their own structure
functions.  Choosing one of the cases considered
by Lazarian and Pogosyan \markcite{lazarian1}(2001), we attempt to determine
the power law spectral indices of the spatial power
spectra of both quantities.  The results suggest that
the different thermal regimes (warm and cool H{\tt I}) are
in fundamentally different regimes of turbulence.

\section{Observations}

The SGPS is a combination of two surveys of the 21-cm line
and continuum emission and absorption from the southern 
Galactic plane ($\delta < -35\arcdeg$, i.e.
$253\arcdeg < l < 357\arcdeg$ at $b=0\arcdeg$).  
The Australia Telescope Compact Array (ATCA)
interferometer of the Australia Telescope
National Facility \footnote{The Australia Telescope National
Facility is funded by the Commonwealth of Australia for 
operation as a National Facility managed by the Commonwealth
Scientific and Industrial Research Organisation.} is used
in its mosaicing mode to make aperture synthesis images, that
are combined with single dish images using the Parkes Telescope
multibeam receiver (Paper 1, Dickey et al. \markcite{jd1}1999).
The latitude coverage is $|b|<1\arcdeg$ for the interferometer
with extension to $|b|<10\arcdeg $ with the single dish only
(resolution $\sim 15\arcmin $).  The resulting spectral line
cubes have uniform sensitivity
to all angular scales down to $\sim 100\arcsec$.
The combination is done using the Miriad task IMMERGE, that provides
the opportunity for careful gain balance between the single
dish and interferometer images, using their respective amplitudes 
in the overlapping portion of the $uv$ plane.  We used the
continuum images (line free) to set this relative gain factor,
the line data are thus not used to scale itself at any stage.
Overall flux calibration is tied to the primary southern 
calibration source, PKS B1934-638, that is assumed to have
flux density of 14.95 Jy at 1380 MHz.

The best studied portion of the SGPS so far is our test region,
which covers $326\arcdeg < l < 333\arcdeg$ and
$0\arcdeg < b < +3\arcdeg$.
Observations of this area were done before the main survey, to 
evaluate the data quality and reduction strategy.  Our basic
spectral line cube of this region has spectral channel spacing
0.824 km s$^{-1}$, pixel size 40\arcsec, and clean beam size (FWHM) 
118\arcsec $\times $ 125\arcsec.  This gives gain G = 41.2 K/Jy,
where G is defined by
\[ G\ =\ \frac{2 k}{\lambda ^2} \times \Omega _{B} \]
where $k$ is Boltzmann's constant, $\lambda$ the wavelength,
and $\Omega_{B}$ is the solid angle of the synthesized beam,
that is 1.3x10$^{-3}$ degrees$^{2}$ or 10.3 pixels in the map plane.
The rms noise is 1.7 K in the spectral line channels, and the dynamic
range is about 250:1 in the continuum.  The area was covered as
a mosaic of 190 pointings with the interferometer, of which each
one was observed 40 times with ``snapshots'' of 30 s integration time
each.  The mosaic was processed with the Miriad program MOSMEM
(Sault et al. \markcite{sault}1996), that deconvolves the effects of missing
$uv$ spacings, and partially recovers the short spacing information
that the interferometer misses.  
 
\begin{figure}
{\bf For figure 1 see the jpeg image associated with this paper.}
\epsscale{0.6}
%\plotone{f1.eps}

\caption{ 
H{\tt I} emission in the SGPS test region.  The upper panel shows
the integral in velocity over the 21-cm spectrum, at each 
pixel in the area.  The grey bar at the top gives the scale,
in units of 10$^{20}$ cm$^{-2}$
($1\times 10^{20}\ = \ 54.85$ K km s$^{-1}$).
Contours are drawn at 50, 100, 150, and 200 $\times 10^{20}\ $cm$^{-2}$.
The light spots appear at the positions of continuum sources that
show strong absorption; they appear negative because the 
continuum has been subtracted from the cube.
The lower panel shows the longitude-velocity diagram of the H{\tt I} 
emission at b=+0\fdg 022.  The stripes near longitude 331\fdg 5
show the effect of absorption toward some faint continuum features,
but generally this latitude has no strong continuum sources, so
the l-v diagram is mostly absorption-free.
Contours are drawn at 5, 30, 55, 80, and
105 K in brightness temperature.
}

\end{figure}

Figure 1 shows the H{\tt I} emission after continuum subtraction.
The upper portion shows the velocity integral of the emission
spectra in each pixel (zero moment map), and the lower portion
shows the emission brightness vs. longitude and velocity along
a line near latitude 0\arcdeg for which there are no bright
continuum sources.  Individual channel maps of the 21-cm 
emission of this region are shown in Paper 2.
The large scale Galactic structures that cross this
longitude range are the Norma Arm, the Scutum-Crux Arm, and
the Sagittarius-Carina Arm.  Following the discussion of 
Paper 2 (Figure 4), the Norma Arm is near the
tangent point, at galactocentric radius about 4.5 kpc and
radial velocity -100 dropping to -60 km s$^{-1}$ (LSR).  We see this
arm tangent to the line of sight, stretching from about 
6 to 11 kpc distance.  The Scutum-Crux arm is at about 3.5 kpc
distance, with radial velocity -50 to -60 km s$^{-1}$.  
The Sagittarius-Carina Arm blends with the local gas at about
-10 to -20 km s$^{-1}$.  The solar
circle on the far side is at about 15 kpc distance.  At that
distance the z-height of the top of our field ($b\simeq +3$\arcdeg)
is about 800 pc, so the scale height of the H{\tt I} layer (nominally
150 pc) comes up only to about 0\fdg 6.  At the Scutum-Crux
arm distance the top of our field is only at 180 pc, so 
at -50 km s$^{-1}$ we see a decrease in H{\tt I} brightness with b 
by about a factor of two to three over our latitude range.

This large scale structure of the inner Galaxy is not strikingly
apparent on the longitude-velocity diagram of Figure 1 for 
three reasons.  The first is that, unlike CO and other tracers
of gas in high density environments, the H{\tt I} 21-cm emission coefficient
depends only on the density and not on the temperature.  Atomic
hydrogen is quite widespread, with a filling factor of 25\% or
more in the inner Galaxy, so the inter-arm regions are not empty of H{\tt I}.
But in the high density environment of the spiral arms, most of
the H{\tt I} is converted to H$_2$, so the 21-cm emissivity no longer
traces the total density of hydrogen (Allen \markcite{allen}2001).
Large scale Galactic structure typically causes only moderate
variations in the H{\tt I} emissivity, that are
reflected in variations of a factor of two or three in the
spectrum of brightness temperature vs. velocity.
The second problem is that large scale departures from
circular rotation can cause as much structure in the 
emission spectra as density variations (Burton \markcite{burton1}1971).  Thus
the details of the $l$-v diagram depend as much on the
velocity structure of the spiral pattern as they do on the
local emissivity variations.  Finally, the third complicating
factor is self-absorption in the 21-cm line.  The optical
depth of the H{\tt I} is on the order of one over roughly 30\% of
the inner Galaxy velocity range (here negative velocities).
A full analysis of this effect is beyond the scope of this
paper, we will present a more extensive discussion of
H{\tt I} absorption and H{\tt I} self-absorption in a forthcoming paper.

\section{Analysis}

To study the SPS of the emission,
we want to look where there is emission only,
to avoid confusion between emission and absorption.
We concentrate on two sample regions where there are no bright 
continuum sources, each 128 pixels square (85.3\arcmin $\times$
85.3\arcmin); region 1 extends from 
(l,b)=(327\fdg 32,+1\fdg 82) to (328\fdg 74,+3\fdg 23), 
region 2 extends from (329\fdg 29,-0\fdg 01) to
(330\fdg 70,+1\fdg 40).  The first is at a latitude high 
enough that even the diffuse continuum is faint, the latter is
near b=0\arcdeg, so it has the longest lines of sight across the outer
galaxy.  These areas are indicated on Figure 2, that shows
the continuum emission from the test region.  The extended 
sources are mostly identified as H{\tt II} regions or supernova 
remnants, as discussed in paper 2.  Here we avoid them and their
effects on the ISM.  Figure 2 shows representative spectra
(taken from the center of each area), with at the bottom of
each an indication of the radial velocity vs. distance $r$ along
the line of sight predicted by cylindrical rotation with the
rotation curve of Fich et al. \markcite{fich}(1989).  The {\bf slope} of this
curve is the velocity gradient, a critical quantity for determining
the line of sight depth of a given velocity interval,
and thus the expected profile shape (Burton \markcite{burton2}1989).  
Along most of the line of sight in the inner Galaxy the
velocity gradient has value $\frac{dv}{dr} = \pm 20 \frac{km s^{-1}}{kpc}$,
except near the subcentral point where it necessarily
goes through zero. 

\begin{figure}
\epsscale{0.6}
\plotone{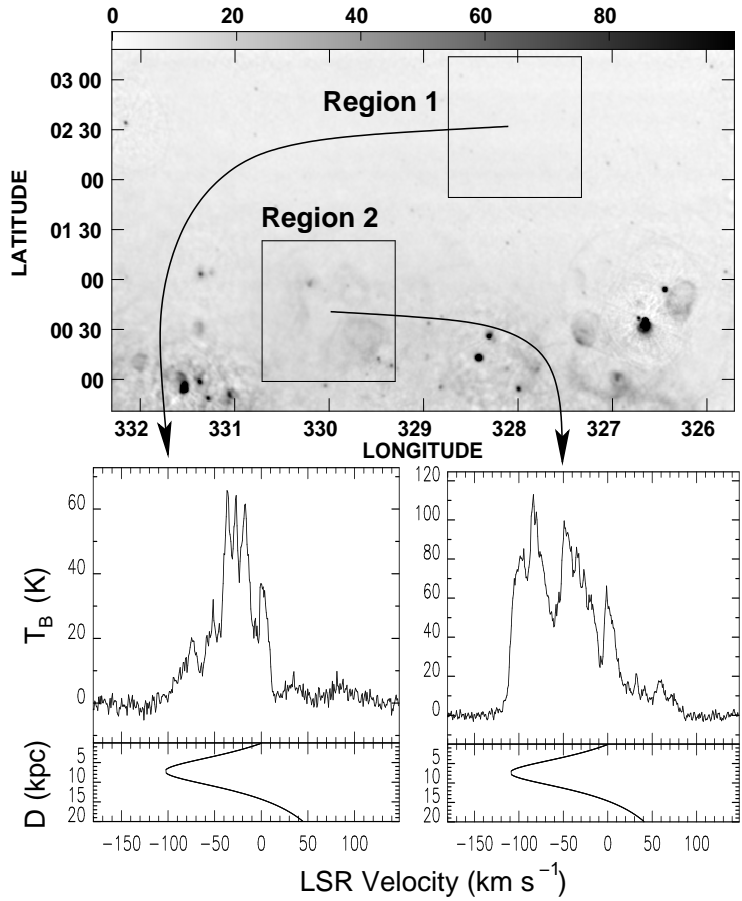}

\caption{ 
Continuum image of the SGPS test field, with the two smaller, 
mostly continuum-free regions indicated.  The grey scale is
in K of brightness temperature.  The H I spectra below are taken
through the central pixel (65,65) of each region.  The scale
below each spectrum shows the kinematic distance using the
rotation curve of Fich et al. (1989).  
}
\end{figure}

For each region, we apodize by multiplying each plane of the cube
by a window function that is unity everywhere except
near the edges of the region where it drops smoothly to zero with
a Gaussian profile (Stanimirovi\'c \markcite{stanimirovic2}1999).  This reduces
the ``edge effects'' in the
transform plane caused by discontinuities at the edge of the 
image when it is replicated in both dimensions 
(e.g. Brault and White \markcite{brault}1971).
Then we Fourier transform, and compute the
{\bf magnitude squared} of the Fourier components.  

The transforms were done using the Miriad task FFT, 
that does the normalization by $\frac{1}{N}$
in the inverse transform, so values on the transform plane 
are typically a factor of $\sqrt{128}$ higher than on the sky plane. 
There are still ``diffraction spikes'' through the center 
of the transform image (pixel 65,65 is the center or ``zero spacing'' 
in the uv plane) due to the narrow width of the edge taper in
the apodizing function, so we blank row 65 and column 65.
Then we segment the uv plane into annuli with radii increasing
logarithmically, and for each annulus we compute the mean
and rms of the values of the transform magnitude squared
for all pixels whose centers lie in the annulus.
The results are shown in Figures 3 through 6.

\begin{figure}
\epsscale{0.6}
\plotone{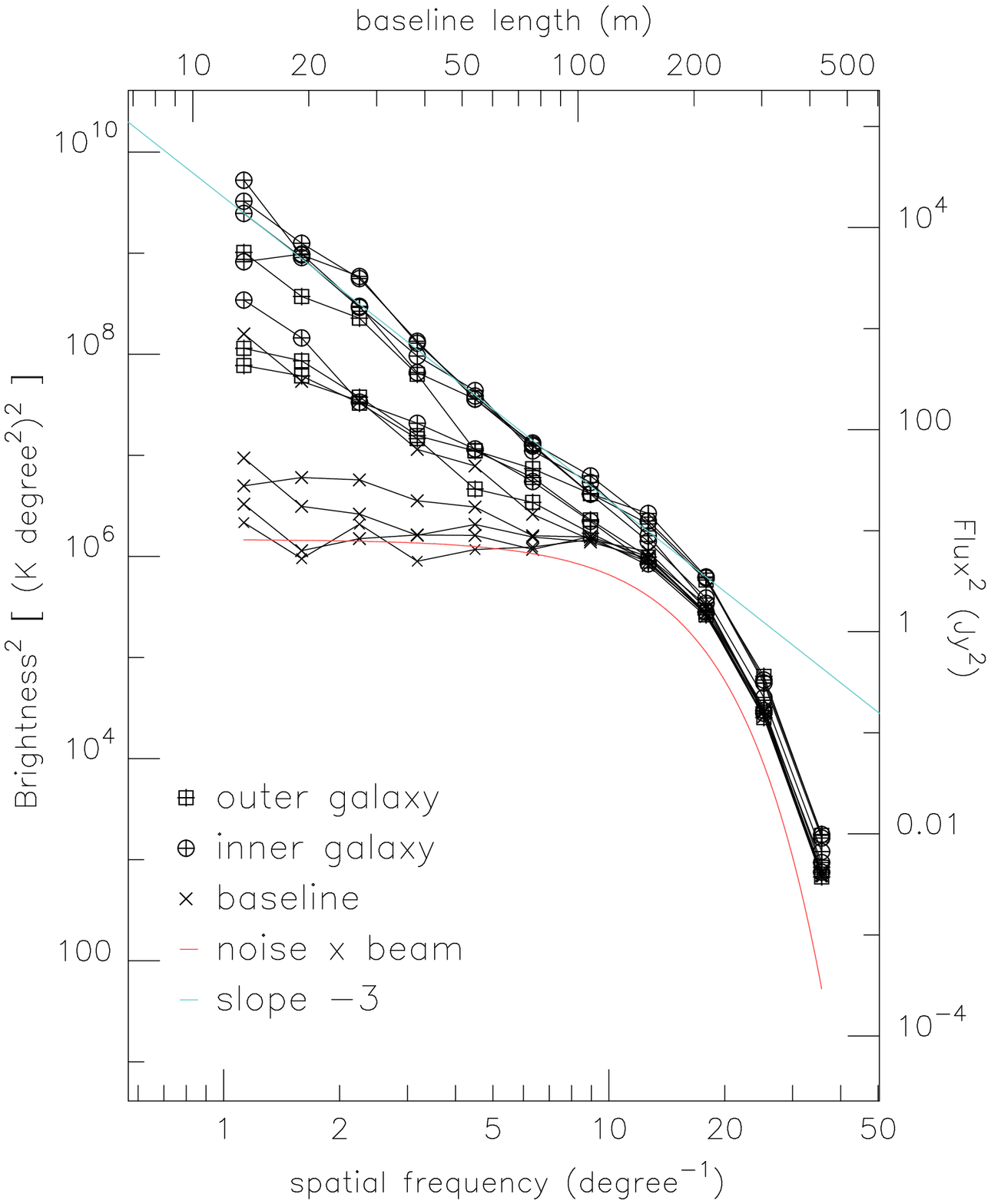}

\caption{ 
The spatial power spectrum of the H{\tt I} emission in Region 1.
The bottom axis scales the spatial frequency, increasing
to the right (smallest angular scales).  The left axis
shows the magnitude squared of the Fourier Transform components
of the H{\tt I} distribution.  The segmented lines with 
symbols show the data for different velocity channels, separated
by 20 km s$^{-1}$, from -140 to +100 km s$^{-1}$ LSR.  Negative
velocities (from the inner Galaxy) and positive velocities
(from the outer galaxy on the far side) are distinguished by 
squares and circles respectively.  Velocities with little or no
emission (``baseline'') are marked with $\times$ symbols.  The lower curve
(red in the electronic edition) shows the prediction for white
noise smoothed with the synthesized beam.  The straight line
(blue in the electronic edition) shows a power law with index -3.
}
\end{figure}

Figure 3 shows the spatial power spectra of the H{\tt I} emission
in region 1.  Each set of points connected by line segments
corresponds to a single velocity channel (0.824 km s$^{-1}$).
We plot only every 24th channel, so the lines on
Figure 3 correspond to velocity channels spaced
by 20 km s$^{-1}$ from -140 to +100 km s$^{-1}$.  These include
several baseline channels on each side of the emission,
five channels that sample primarily the inner Galaxy (at negative
velocities), and four that sample the outer galaxy (positive 
velocities), indicated by different symbols on Figure 3.  

Figure 3 has different units on its different axes.
The bottom axis is the spatial frequency, that has units of
inverse angle (here degrees$^{-1}$).  This shows the radius
on the uv plane of the middle of each annular bin.  Corresponding to 
a given spatial frequency (number of cycles per degree) is a baseline length
of the interferometer; these lengths are shown on the top scale.
The original image has units of brightness temperature (K), so
its transform has units of brightness temperature per pixel 
area, or K degree$^{+2}$.  Since the images are 128 x 128 pixels at 40\arcsec,
the pixel size on the conjugate
($uv$) plane is 0.703 degrees$^{-1}$ and the pixel area is
0.494 degrees$^{-2}$.  Finally we scale the magnitude squared
of the transform components by the
pixel size squared, i.e. by dividing by (0.703 deg$^{-1}$)$^{4}$
to get units of K$^2$ degree$^{+4}$.  These units are indicated
on the left hand scale of Figure 3.

We can alternatively convert the original image to units of Jy per beam by 
dividing by 41 K/Jy (the gain of the synthesized beam, G, defined above).
Dividing the image in units of Jy per beam by the beam area in pixels
gives units of Jy per pixel.  Then the units on the transform
plane are simply Jy.  
The corresponding squared magnitudes have units Jy$^2$, as indicated
on the right hand scale of Figure 3.

\begin{figure}
\epsscale{0.6}
\plotone{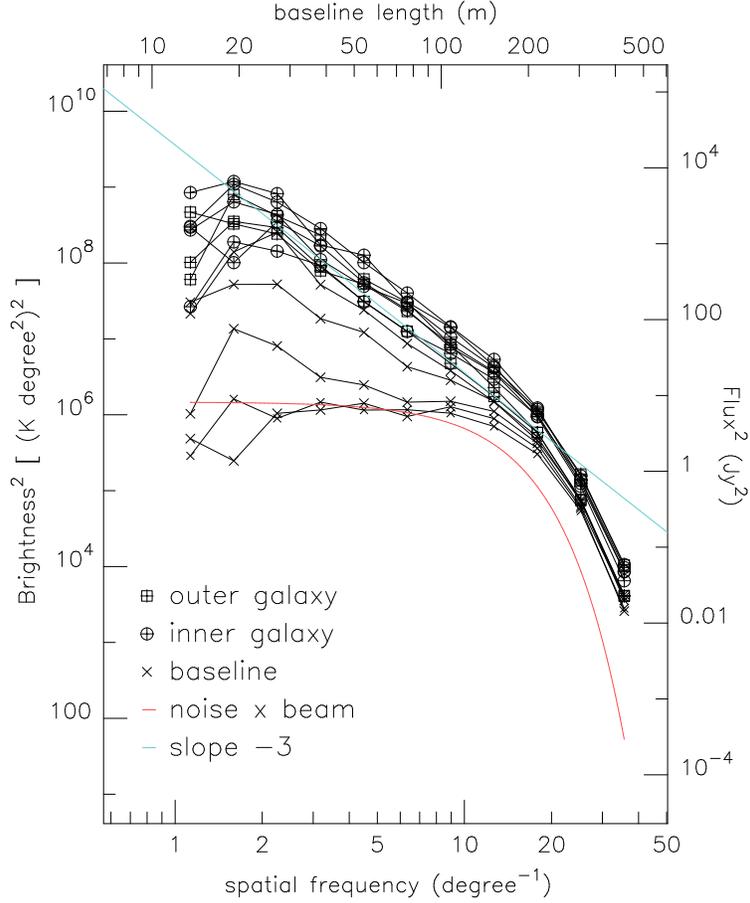}

\caption{ 
The spatial power spectrum of the H{\tt I} emission in Region 2,
with axes similar to those on Figure 3.
This case uses the interferometer data alone, without combining
with the single dish images, hence the turnover at low spatial
frequencies (angles larger than about 20\arcmin are not well
sampled by the interferometer).  The straight line (blue
on the electronic edition) is {\bf the same} as on Figure
3, that shows that the H{\tt I} emission is generally stronger at
the lower latitudes, particularly at high negative velocities
(near the subcentral point) and at positive velocities in the
outer galaxy on the far side of the solar circle.  This is
an effect of the lower latitude of this region.
}
\end{figure}

On Figures 3 and 4 the baseline channels show flat spectra, 
consistent with radiometer noise that is spatially ``white'',
up to the inverse of the synthesized beam size
($\simeq 120$\arcsec FWHM).  The prediction
for a Gaussian with FWHM = $\frac{2}{120\arcsec}$ and amplitude 
equal to the square of the two sigma noise level is indicated on
Figures 3 and 4.  Channels for which the region
is filled with emission show the familiar power law
SPS, with spectral index very close
to -3 out to the resolution limit.  Figure 4 uses only
the ATCA data for region 2, while Figure 3 uses the combined
Parkes plus ATCA images for region 1.  The interferometer data
alone show a cutoff at large angular sizes (the last 
column of points
on the left side of Figure 4) that can only be measured
with the single dish telescope.  The smoothness of the spatial
power spectra down to this smallest spatial frequency in Figure 3 
demonstrates the quality of the calibration between the interferometer
and single dish data.  A few channels on Figures 3 and 4
show power law slopes flatter than -3, for example v=+80 km s$^{-1}$
in region 1 shows slope of -1.5.  Investigation of the images shows
that the emission at this velocity is concentrated in a very 
small clump that is presumably at great distance in the outer
galaxy (its nominal kinematic distance is $37$ kpc).  Since this emission
covers an area much smaller than the area of region 1, the spatial
power spectrum is dominated by the {\bf boundaries} of the cloud, rather
than its internal structure. This inevitably flattens the slope
of the spectrum, since the lower spatial frequencies cannot be
as strong for a small cloud as for emission that fills the region.
In the limit of a very small cloud, barely resolved by the synthesized
beam, the SPS would be flat, similar to the noise in the baseline channels
on Figures 3 and 4.  The different slope in this case
is not evidence for different
statistics of the {\bf internal structure} of the H{\tt I} cloud,
that could only be measured in this way over much smaller
areas entirely within the cloud's boundaries.

\begin{figure}
\epsscale{0.6}
\plotone{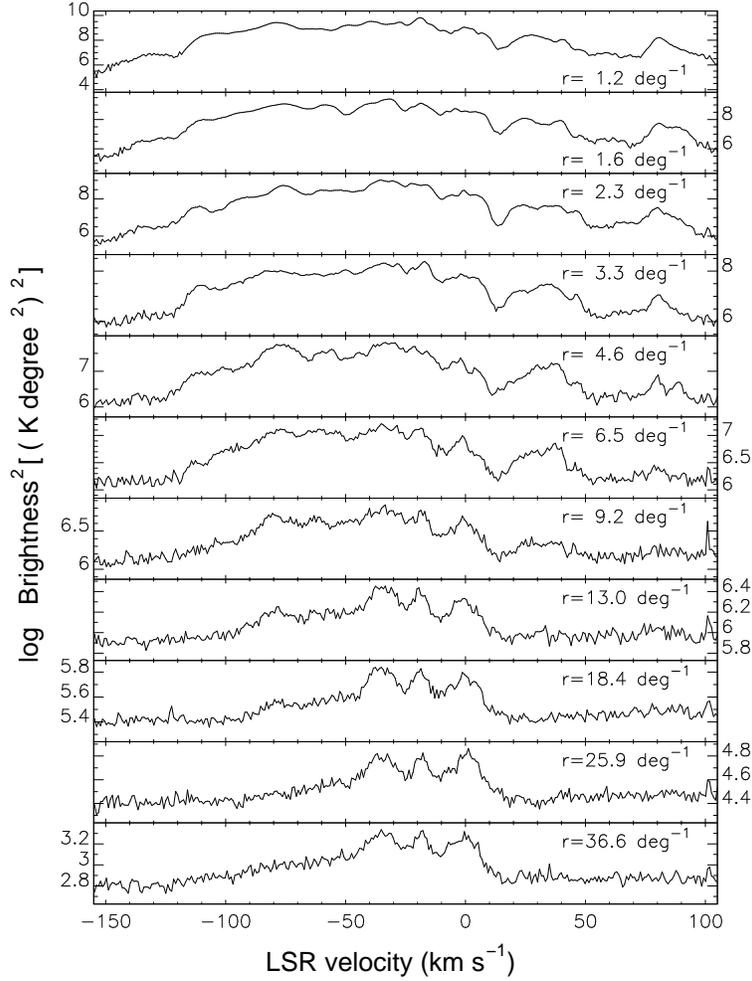}
\caption{ 
Velocity profiles of the SPS amplitude, at
various $uv$ radii.  Each panel corresponds to a different column
of points on Figure 3, the top panel corresponding to the
leftmost column, the bottom to the rightmost column.  Whereas on
Figure 3 only a few velocity channels are shown, here the entire
velocity range is displayed.  Note that the vertical scale changes
drastically from the top profile to the bottom, as indicated by
the values on the scales, which alternate from side to side.
For this reason the noise is much more noticable on the lower
profiles compared with the upper ones.
}
\end{figure}

%\begin{figure}
%\epsscale{0.6}
%\plotone{f6.eps}
%\caption{ 
%}
%\end{figure}

Figure 5 shows velocity spectra of the 
amplitude squared of the Fourier components.  Each panel 
shows the average over one of the annuli on the $uv$ plane,
starting with radius 1.2 degree$^{-1}$  and going to 
37 degree$^{-1}$ in steps of 0.15 in
the log$_{10}$ of the radius (i.e. factors of $\sqrt{2}$ in spatial
frequency, as on Figures 3 and 4).
The lowest spatial frequencies show broad, very smooth spectra,
whose most significant feature is a bump at high positive velocities
($\sim +80$ km s$^{-1}$) on Figure 3.  This corresponds to the
small cloud mentioned above.  At higher spatial frequencies the
spectra show more structure in velocity, but they are still
smoother than any single-pixel emission spectrum from
the region.

\subsection{Velocity Averaging}

The results presented in Figs. 3 and 4 show that the H{\tt I} spatial
power spectrum for both regions 1 and 2 can be fitted with a power law,
over a range of angular scales from 3 arcmin to 1 degree (corresponding
to length scales of 6 to 130 pc at the tangent point distance of 7.4 kpc).
As region 1 is located at $b \approx 2$ degrees
($z \simeq 250$ pc at the tangent point) and does not contain
significant continuum emission, the effects of
absorption and H{\tt I} self-absorption 
are not severe, and the 21-cm line
intensity is generally directly proportional to
the H{\tt I} column density. We can therefore test
predictions of the turbulence theory
described in Lazarian \& Pogosyan
(\markcite{lazarian1}2000, hereafter LP)
on this region.  This theory provides a way
to relate the observed 2-D SPS of the H{\tt I} brightness
to the underlying 3-D statistics of 
both the density and the velocity distributions. 

The simplest theoretical situation is the case of
density fluctuations alone, 
with no small scale structure in the velocity field.
In this case, a narrow section through the medium should show
a SPS slope that differs from that of a deep sample. 
In the asymptotic limit of thin and thick slices described
by LP (section 3), and assuming steep spectra (slope $n$ less 
than -1) the difference in slope is just one unit,
i.e. 

\[ P({\bf k}) \ \propto \ \begin{cases} 
|k|^{n+1} \ \ \ \ \ \  \textrm{thin} \\
|k|^{n} \ \ \ \ \ \ \ \textrm{thick}
\end{cases} \]

\noindent where $P$ is the observed SPS and $k$ is the
spatial frequency.    

We can use velocity smoothing to effectively change the
width of the slice, since the velocity gradient translates
velocity width to line of sight depth as described above.
To search for a change of slope, we average 
the line data over broad velocity widths in the cube
for Region 1, and then recompute the SPS following
the same method described above.  The results are shown on
Figures 6 - 8.  Figure 6 is similar to figures 3 and 4,
but with the spectral line data averaged over 24 channels
before the Fourier transform (bandwidth 20 km s$^{-1}$).
Based on the velocity gradient of 20 km s$^{-1}$ kpc$^{-1}$,
averaging over this velocity width increases the line of
sight depth of the samples from $\sim$0.4 to 1 kpc.  The
SPS does indeed steepen, as predicted, from
$\sim$ -3.1 to $\sim$ -4.0 in the logarithmic slope.
This range includes the prediction of the simple
Kolmogorov theory, $n = -\frac{11}{3}$.

\begin{figure}
\epsscale{0.6}
\plotone{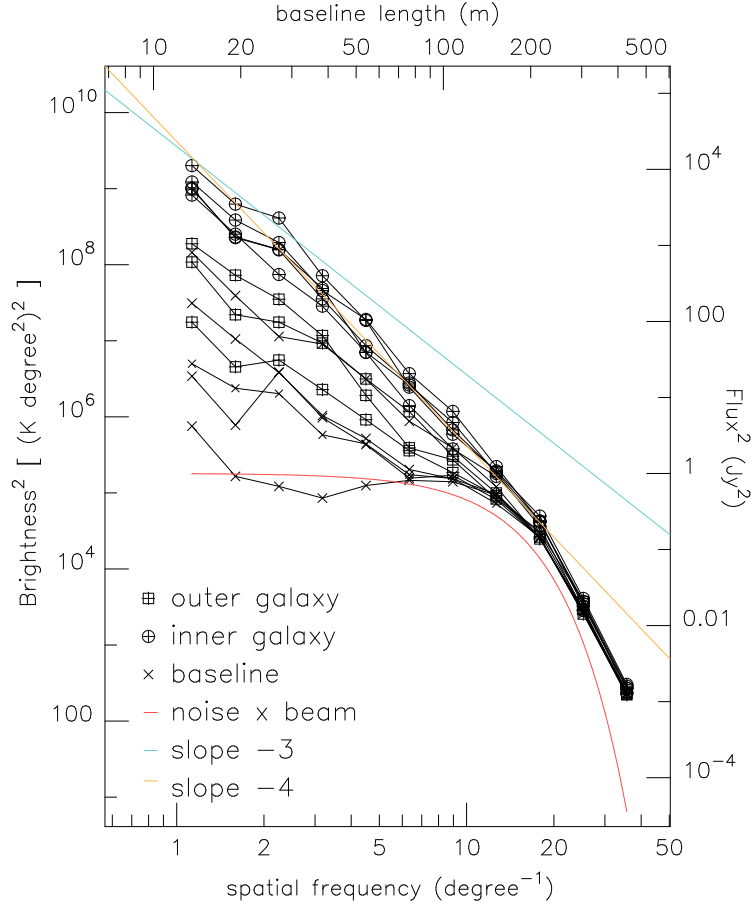}
\caption{ 
The SPS of the {\bf velocity averaged} H{\tt I} emission
of region 1.  This is similar to Figure 3, except that instead of 
analysing single-channel data separated by 20 km s$^{-1}$, we first
average over 20 km s$^{-1}$ velocity widths (24 channels) and then
Fourier transform.  The line with slope -3 (blue on the electronic
edition) is the same as that on Figure 3.  The steeper line 
(gold on the electronic edition) has slope -4.  For the inner
galaxy velocities, where the region is mostly covered with H{\tt I} 
emission, the slope has changed from -3 to -4 due to the
velocity averaging.
}
\end{figure}

We study this transition
from thin to thick sections through the gas, by averaging over progressively
broader bandwidths from a single 0.8 km s$^{-1}$ velocity
channel, to three (2.5 km s$^{-1}$), six
(4.9 km s$^{-1}$) and so on up to the 24 channel averages shown on Figure 6.
We then fit straight lines to the spatial power spectra over
the interval from 2 to 8 deg$^{-1}$ for each new velocity average.  Results
are shown on Figure 7 for representative velocity channels spaced by
30 km s$^{-1}$.  Here
we include three channels from the inner Galaxy (at -20, -50, and -80
km s$^{-1}$), two from the outer galaxy (+40 and +80 km s$^{-1}$), and one 
baseline channel that is mostly noise (-140 km s$^{-1}$).  They all 
show some steepening, although the baseline channel fluctuates
between slope 0 and slope -1.5.  At velocity +70 the slope steepens
from -0.5 to -2, but this is probably simply due to the small
covering factor of this very distant gas, so that the broader velocity
averaging includes many separate, unrelated structures in such a way
as to increase the covering factor of the H{\tt I}.
The inner Galaxy channels are the most consistent and they
best match the prediction of the turbulence model.  Each one steepens 
smoothly (from slope -2.5 to -3.6 at -50 km s$^{-1}$, from -3.3 to -4.1
at -20 km s$^{-1}$, and -3.1 to -3.6 at -80 km s$^{-1}$).  The last, -80 km s$^{-1}$,
seems to flatten at the broadest bandwidths, that may be due to the 
velocity gradient's sharp drop to zero at the terminal velocity ($\sim$
-100 km s$^{-1}$ at this longitude).  

\begin{figure}
\epsscale{0.6}
\plotone{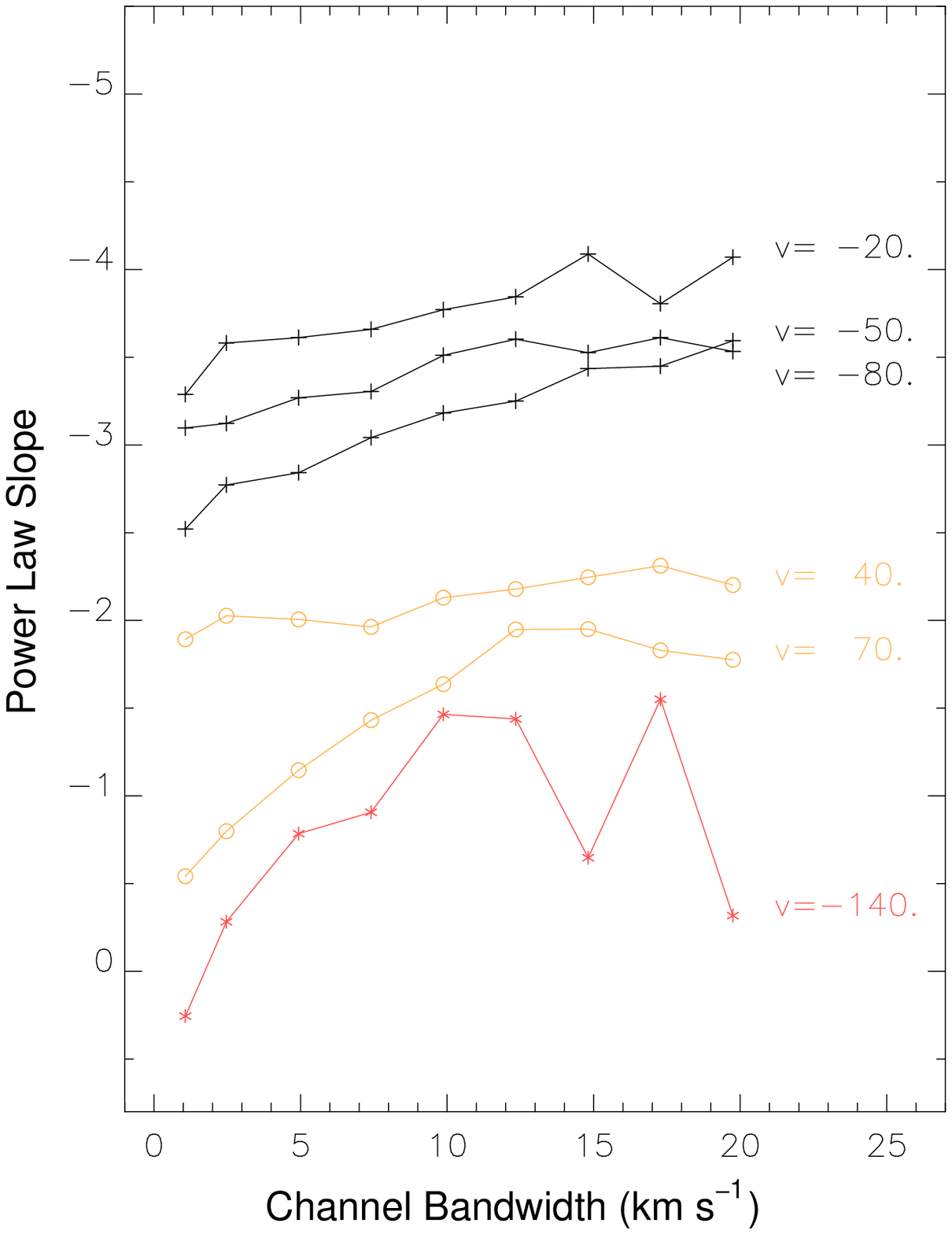}
\vspace{-1in}
\caption{ 
Slopes of the spatial power spectra for various velocity averaging.
Each point represents a least squares fit to the logarithmic slope
of the spatial power spectra (as shown on Figures 3, 4, and 6) for specific
velocity channels.  The cube is first averaged in velocity over 
progressively broader bandwidths (scaled on the bottom), thus each
successive point from left to right represents an average, of one,
three, six, etc. up to 24 velocity channels (20 km s$^{-1}$ width).
The three upper curves, marked with crosses (black in the electronic
version) are for the inner Galaxy, centered at velocities of -20,
-50, and -80 km s$^{-1}$.  The bottom curve (red in the electronic edition)
is mostly noise, it is a baseline channel centered at -140 km s$^{-1}$
well beyond the terminal velocity.  The middle curves, marked with
circles (gold in the electronic edition), are at positive
velocities that cover distant gas, well
beyond the solar circle on the far side of the galaxy in the lower
halo, which does not fill the field (above the noise).
Typical error bars (representing both noise and the dispersion
of the ensemble as measured from independent velocity channels)
are $\pm 0.25$ on the upper curves.
}
\end{figure}

Averaging more than about 20 km s$^{-1}$ does
not significantly steepen the slope.  This is shown on the
upper panel of Figure 8, that
plots {\bf all} data for the inner Galaxy for a broad range of
averaging widths (to 55 km s$^{-1}$).  To make Figure 8 we compute
the slope for every modified velocity channel, and then find
the mean and standard deviation of the slopes.  The error
bars indicate $\pm \frac{\sigma}{\sqrt{s}}$  around the mean
of the slopes, for the sample (size $s$) of all modified
velocity channels with
centers in the range -15 km s$^{-1}$ to -80 km s$^{-1}$.

\begin{figure}
\epsscale{0.5}
\plotone{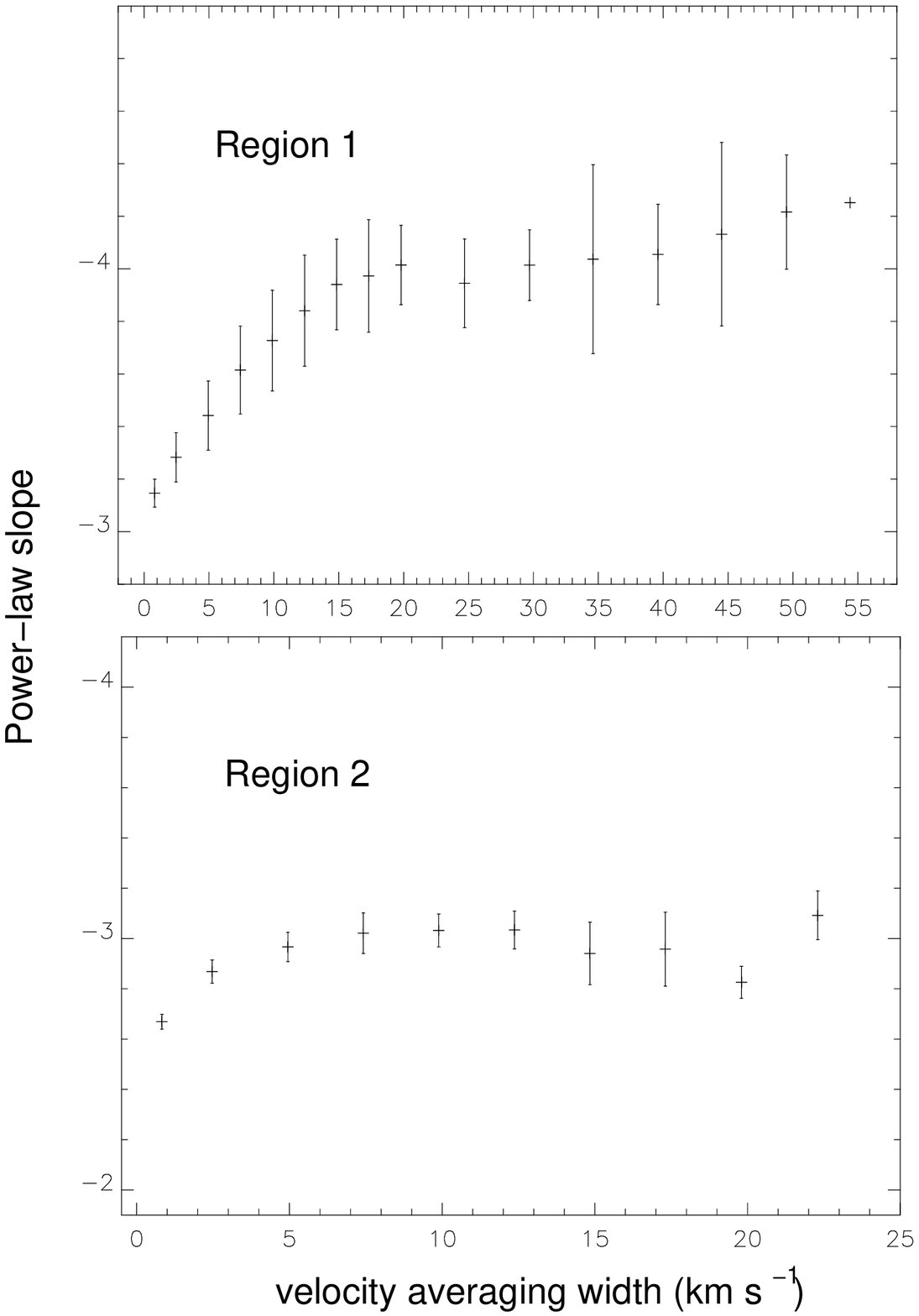}
%\vspace{-1in}
\caption{
Power law slope vs. velocity averaging width.  In this case
we consider {\bf all} channels with center velocity in the range 
-15 km s$^{-1}$ to -80 km s$^{-1}$ for each averaging width, and compute
the mean and standard deviation of the slopes of the power spectra
in each case.  Results
for both regions are shown.  Region 2 is at such low latitude
that self-absorption in the hydrogen is significant; this
may mostly cancel the velocity averaging effect in the power
law slope.
}
\end{figure}

Including the turbulence-induced fluctuations of the {\bf velocity field}
of the gas as well as density fluctuations changes the meaning of
the slope.  Following LP (section 4.2) we find that the power law 
slope of the SPS for the thin and thick slices becomes
$(-3 - \frac{m}{2})$ in the thick case and $(-3 + \frac{m}{2})$
in the thin case,
where $m$ is the power law index of the velocity structure function.
The Kolmogorov case has $m=\frac{2}{3}$, that would give -2.67
and -3.33 respectively for the slopes in the thin and thick cases.
If we try to force the data from region 1 to match the prediction
for the velocity
dominated regime, we find a steeper velocity spectral index
($m=0.9$) but {\bf not centered on -3}.  This suggests that the
velocity term is not strong enough to dominate the density
term in the calculation of the SPS slope for region 1,
so there we can measure $n$ but not $m$.

Region 2, at much lower latitudes, shows very different behavior.
The center of the region is at $b=+0\fdg 7$, that translates to
$z = 90$ pc at the tangent point.  This is less than the scale 
height of {\bf cool} H{\tt I} (roughly 120 pc), so the line of sight is
rich in cool gas, as shown by the many deep absorption lines
toward continuum sources in this area.  This effects the LP
analysis of the SPS in two ways.  First, self-absorption in
the H{\tt I} causes the brightness temperature to saturate, so that
it no longer traces the column density of the gas.  Second,
the intrinsic linewidth of the emission changes from about
16 km s$^{-1}$ to about 1.6 km s$^{-1}$, since the thermal 
velocity decreases with the temperature, that is
about 6000 K in the warm H{\tt I} but typically 60 K or less in
the cool gas.  
The thermal linewidth is given by $\Delta v (FWHM) \ 
= \ 2.35 \sigma_v$ where $\frac{\sigma _{v}}{km\ s^{-1}} \ = \
\sqrt{\frac{T}{121\ K}}$.  
This could have a dramatic effect on the 
SPS, as discussed by LP (section 4.3), in that a given 
velocity width may be effectively thick for warm gas, but
thin for cool gas.  LP predict that changing the velocity 
resolution will change the SPS slope only so long as the
velocity resolution is less than the thermal velocity width.
The lower panel of figure 8 shows that the steepening of 
the SPS slope in region 2 stops at velocity width of about
6 km s$^{-1}$.  This is rather broad for the {\bf thermal}
width of the cool phase gas, 
however 6 km s$^{-1}$ matches
quite well the random velocity of the cool H{\tt I} clouds
(Belfort and Crovisier \markcite{Belfort}1984).  Thus
the implication of figure 8 is that at low $z$, where
the cool phase of the H{\tt I} dominates, the turbulent 
{\bf velocity} field dominates the emission fluctuations,
whereas at higher $z$, where there is little cool
H{\tt I} relative to warm, the {\bf density} fluctuations dominate.

The numbers indicated for region 2 on figure 8 are significantly
lower than for region 1; the power law index varies from about -2.65
to about -3.1.  The much smaller increase in slope with increasing 
velocity width in this case suggests that we are now in the regime
dominated by velocity fluctuations, rather than density fluctuations
as in region 1.  The narrow band slope of -2.65 is in good agreement
with the Kolmogorov prediction for the velocity dominated regime
(-2.67), but the steeper slope of -3.1 for the broader velocity
slices does not match the Kolmogorov prediction of -3.33, based
on $m=\frac{2}{3}$.  Taken at face value, our slope change suggests
that $m \simeq 0.2$, that is a much shallower slope than in the
Kolmogorov case, implying more energy on small scales than in the
classical theory.
These numbers are remarkably similar to the power-law index of -2.8
found for {\bf molecular clouds}, using the optically thick lines
of $^{12}$CO and $^{13}$CO as tracers (St\"utzki \markcite{stutzki}et
al. 1998; Bensch \markcite{bensch}et al. 2001).
This points out that optically thick gas may show a significantly different  
SPS slope from the optically thin case.
Further modifications of the LP
theory are needed in order to relate the 2-D and
3-D properties of gas in the optically
thick regime. 

Recent measurements of the SPS of the H{\tt I} in the SMC
(Stanimirovic \& Lazarian 2001, in 
preparation) suggest that in that environment the velocity fluctuations
dominate the density fluctuations even for the warm medium. The
SPS in the SMC shows a gradual steepening of the power-law
slope, from $-2.8$ to $-3.3$, with velocity averaging width.  This is
quite close to what would be expected for Kolmogorov turbulence ($m=\frac{2}{3}$),
for an optically thin tracer for which velocity fluctuations dominate. 
Comparison with the spectra obtained for region 1 suggests that 
while the SMC has more power on smaller scales, the energy cascades faster 
in the Milky Way than in the SMC.  

\section{Conclusions}

This is the first study of the Galactic H{\tt I} spatial power 
spectrum to make use of the technique of large-scale
mosaicing that was pioneered by Staveley-Smith et al.
\markcite{LSS1}(1995) in their
mapping of the SMC.  This technique allows
analysis of many interferometer primary beam areas together,
at the full resolution of the maximum interferometer baseline.
Combining the single dish and interferometer data gives us
an effective range of spatial frequencies of more than a factor
of 30.  In addition, the mosaicing technique removes most of
the effect of the attenuation
due to the primary beam of the interferometer, that was 
a complication in earlier single-field interferometer
studies of the H{\tt I} in the Milky Way.
One motivation for this analysis was to confirm the data quality
of the SGPS, and particularly the relative calibration of the
brightness on large and small angular scales.  The smoothness of
the power laws at low spatial frequencies 
on Figure 3 demonstrates that the SGPS images
do not suffer from systematic bias between the single dish
and interferometer data, or from missing baselines on the 
$uv$ plane.

The results for the SPS slopes agree with the
general prediction of turbulence theory that the slope should steepen
when we make the transition from thin to thick slices
(LP equation 28).
Looking in more detail at the change of slope vs. velocity averaging
in section 3.1 above, we find that the data for region 1 suggest that
in the warm phase H{\tt I} the SPS is dominated by 
{\bf density} fluctuations, that match the Kolmogorov
prediction quite well ($n=-4$ is observed, vs. 
the predicted $n=-\frac{11}{3}$).  Apparently velocity
fluctuations have little effect in this regime. 
However at lower latitudes in region 2 we find that velocity
fluctuations dominate, with the very shallow spectral index
of $m=0.2$.  This interpretation may require further theoretical 
analysis taking account of the optical depth of the cool gas.

The larger significance of the change of slope with velocity width
is that it provides
support to the overall interpretation that the small scale structure
of the ISM is governed by a turbulence cascade.  This has
long been the paradigm for interpretation of the structure 
function of the {\bf ionized} gas, but it has only recently been
extended to cover the structure of the neutral medium as
well.  There is some evidence that extending the SPS
down to the very small angular scales sampled
by VLBI observations of 21-cm absorption variations (Faison et
al. \markcite{faison}1998) shows that even this ``tiny scale structure'' 
(Heiles \markcite{heiles}1996) is consistent with the
same cascade (Deshpande \markcite{desh}2000).

As a disclaimer we must point out that the results presented here,
are derived only from small areas that may not be representative
of the Milky Way disk as a whole.  It is very likely that different
regions of the galaxy have very different characteristics that 
may be reflected in differences in the statistics of their small scale 
density and velocity fields.  The differences between the two regions
studied here may not nearly span the range of possibilities for the
real ISM.  Only a comprehensive study of the data from the entire
SGPS area will show how diverse the HI spatial power spectrum may
be.  On the theoretical side, questions remain about how the
energy of the turbulence cascades from large to small scales,
whether some is lost by dissipation at intermediate scales
since the medium almost certainly is compressible, and how the
irregularities in the density and velocity fields are coupled.
These fundamental questions are beyond the scope of this study,
but we can hardly begin to understand interstellar turbulence 
until we can answer them.

Observationally, the results presented here for the spatial
fluctuations of the 21-cm {\bf emission} are critical for the
measurement and interpretation 
of the 21-cm {\bf absorption} in the SGPS.  This will
be the topic of a forthcoming paper.

\section{Acknowledgements}

We are very grateful to the ATNF staff for their assistance and
encouragement at all stages of this project, and in particular to
Robin Wark and John Reynolds for their support.  
We are grateful to Avinash Deshpande, David Green, 
Itzhak Goldman and the anonymous referee for suggestions and comments
on the manuscript.  B.M.G. acknowledges the
support of NASA through Hubble Fellowship grant HST-HF-01107.01-A awarded
by the Space Telescope Science Institute, which is operated by the
Association of Universities for Research in Astronomy, Inc., for 
NASA under contract NAS 5--26555.  N.M. M.-G. is supported by NASA
Graduate Student Researchers Program GSRP) Fellowship NGT 5-50250. 
This research was supported in part by NSF grant AST 97-32695 to the 
University of Minnesota.

\end{document}